\documentclass[conference]{IEEEtran}
\IEEEoverridecommandlockouts
\usepackage{cite}
\usepackage{amsmath,amssymb,amsfonts}
\usepackage{algorithmic}
\usepackage{graphicx}
\usepackage{textcomp}
\usepackage{xcolor}
\def\BibTeX{{\rm B\kern-.05em{\sc i\kern-.025em b}\kern-.08em
    T\kern-.1667em\lower.7ex\hbox{E}\kern-.125emX}}

\newlength\mylength
\begin{document}

\title{Automated Taxi Booking Operations for Autonomous Vehicles\\
}

\author{\IEEEauthorblockN{Linh Van Ma, Shoaib Azam, Farzeen Munir, Moongu Jeon}
\IEEEauthorblockA{\textit{
School of Electrical Engineering and Computer Science} \\
\textit{Gwanju Institute of Science and Technology, Gwangju 61005, Korea}\\
\{linhvanma, shoaibazam, farzeen.munir, mgjeon\}@gist.ac.kr}
\and
\IEEEauthorblockN{Jinho Choi}
\IEEEauthorblockA{\textit{School of Information Technology} \\
\textit{Deakin University, Geelong, VIC, Australia}\\
jinho.choi@deakin.edu.au}
}

\maketitle
%This research paper presents a part of our on-going research on autonomous vehicle (AV) driving system. We completely build a taxi booking system which assists autonomous vehicles to pick up customers. In conventional taxi booking system, customers and drivers use two different mobile applications connected to a server. Customers use a customer's mobile application to call a taxi, and drivers use a driver's application to accept a booking request from a customer. In our system, the driver's application is a built-in application in an autonomous vehicle. Apart from customer and AV, we build a server to monitor customers and AVs. It also supports inter-communication between a customer and an AV once AV decided to pick up a customer.
\begin{abstract}
In a conventional taxi booking system, all taxi operations are mostly done by a decision made by drivers which is hard to implement in unmanned vehicles. To address this challenge, we introduce a taxi booking system which assists autonomous vehicles to pick up customers. The system can allocate an autonomous vehicle (AV) as well as plan service trips for a customer request. We use our own AV to serve a customer who uses a mobile application to make his taxi request. Apart from customer and AV, we build a server to monitor customers and AVs. It also supports inter-communication between a customer and an AV once AV decided to pick up a customer.
\end{abstract}


\section{Introduction}
Taxi booking applications are widely used nowadays with the support of transportation network companies, such as Uber, Lyft, Inc., Kakao T (a Korean transportation service apps). Recently, many researchers have been carrying out the model of taxi booking system using AVs \cite{dandl2018comparing,dandl2017microsimulation}. Hyland \cite{hyland2018real} thoroughly presented research on shared-use autonomous vehicle mobility services. The author addressed a future paradigm of shared-use autonomous vehicle services by modeling, controlling and then simulating the real-time operation of AV services. Dandl \cite{dandl2018booking} did another research on booking processes for AVs. The authors modeled a booking system and presented optimal methods to pick up a customer based on his origin with many constraints such as price per ride or timing. In early 2019, Todd \cite{litman2017autonomous} addressed in the report of autonomous vehicle implementation predictions that there is still a long way to reach the final stage of fully automated AVs. On the basis of a predictable development pattern (S-curve), the author specified that we have just completed around 20\% of AV development to reach the saturation of AVs market. The author also stated that half of all new vehicles will be autonomous around 2040, and half of the vehicles will be autonomous around 2050.

In this article, we present our taxi booking system as follows. First, we build an AV \cite{munir2018autonomous,mlvazam}, which provides a taxi service, to automatically pick up a customer based on his/her request. Secondly, a customer's application can be run on Android. It allows a customer to push a taxi booking request to a server. Thirdly, an intermediate server, which can run and deploy on Amazon Web Service (AWS), listens for taxi requests from customers and updates location information of customers and AVs in real-time. This location information is employed to find the best AV for a request from a customer. In short, our paper addresses a practical implementation of automated taxi booking operations, and use our own built autonomous car to demonstrate our proposed system.

The rest of this article is organized as follows. Section \ref{secII} provides a literature review of AV booking systems. Section \ref{secIII} describes our booking system overview. We present the detail implementation and demonstrate our booking system in Section \ref{secIV}. Finally, we conclude our work in Section \ref{secV}.

\section{Related Works}
\label{secII}
Self-driving cars, the heart of the autonomous taxi booking system, have been taking much attention from many stakeholders, such as Alphabet Inc. with Waymo, Uber's Advanced Technologies Group, Tesla Inc., and many more motor manufacturers (General Motors, Audi, Volkswagen, Ford). In August 2016, NuTonomy, a start-up company, was first introduced autonomous mobile robots (Robo-taxi) \cite{ackerman2017hail} and self-driving cars to the public. This Robo-taxi is now being testing and developing mainly in Singapore. The closest to our research is Waymo One \cite{hawkins2017waymo,davies2017waymo} which is now offering fully self-driving service in the Phoenix metropolitan area of the United State of America. Waymo autonomous taxi system has one mobile application that assists users to select the best pickup location based on the current location of a customer. The application also shows travel duration as well as a planning route before requesting a ride. To the best of our knowledge, we did not find any technical report, as well as the detail implementation of Waymo One, published yet. However, we can infer that they would have a server that assists the car to select the best pickup location. The server might also plan an optimal route between the origin and destination specified by a customer. In our system, we use the Global Position System (GPS) location as a standard coordinate to share location information among autonomous vehicles and customers. We also build our own vector map \cite{javanmardi2017autonomous} to search and plan an optimal route for our AV.
%https://www.nutonomy.com/
%\footnote{https://waymo.com/}
% as shown in Fig. \ref{fig01} (the first image from left to right)

%\begin{figure}
%\centerline{\includegraphics[width=0.47\textwidth]{figs/waymoone.png}}
%\caption{An illustration of cab booking system for AV of Waymo One.}
%\label{fig01}
%\end{figure}

\section{Taxi Booking System Overview}
\label{secIII}
In our taxi booking system, we use autonomous vehicles to serve customers. A server application collects GPS location of both AV and customers to assign a vehicle base on a customer ride request. A mobile application allows a customer to select his origin and destination for a ride and sends his request to the server.

\begin{figure}
\centerline{\includegraphics[width=0.48\textwidth]{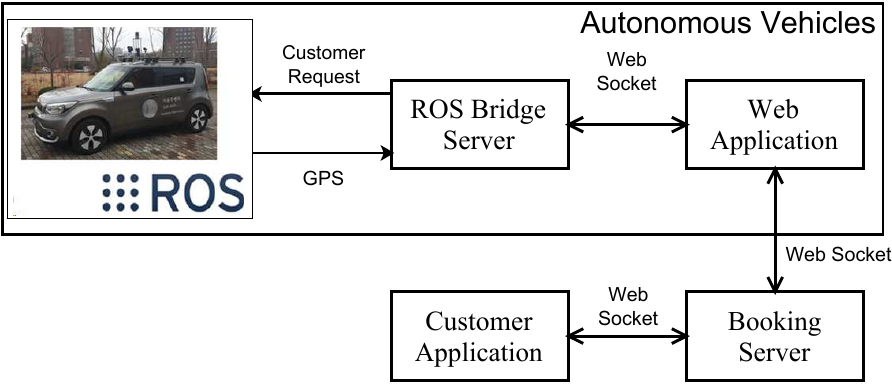}}
\caption{Booking system framework from autonomous vehicle perspective.}
\label{fig11}
\end{figure}

Fig. \ref{fig11} depicts our booking system framework from our AV perspective. More details, we have used our autonomous car which uses Robot Operating System (ROS) as a middleware for communication between different components as well as controls our autonomous car. ROS can be simply defined as a collection of software for robot software development. ROS itself cannot communicate directly with the server application. Hence, we utilize a ROS bridge server to retrieve the GPS of our AV. We also create a web application is an intermediate between ROS bridge server and the server application. It enables ROS bridge server to push our AV current location and retrieves customer's pickup location in real-time from the server application. The web application also visualizes our AV location and the pickup location of a customer on Google map.

\begin{figure}
\centerline{\includegraphics[width=0.48\textwidth]{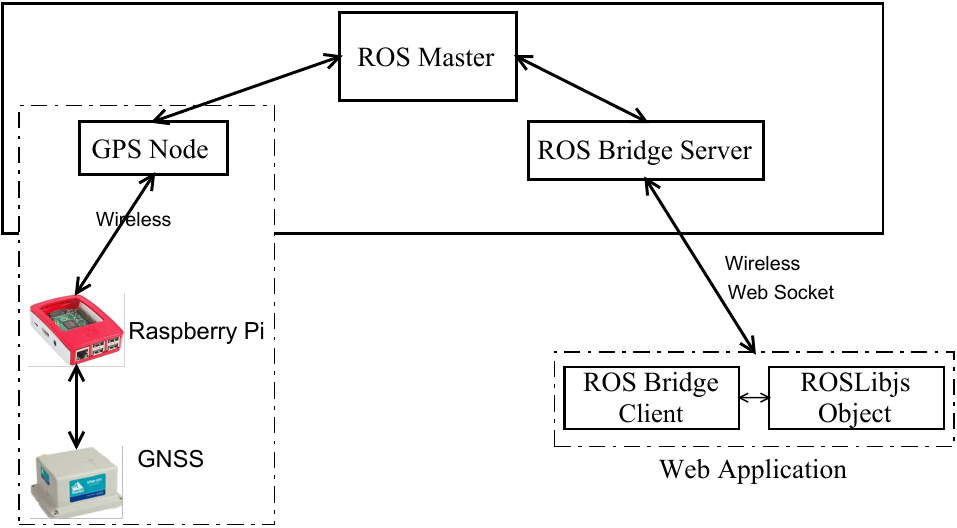}}
\caption{Communication schema between ROS and non-ROS system on autonomous vehicles.}
\label{fig12}
\end{figure}

Fig. \ref{fig12} illustrates the communication schema between ROS and non-ROS system of our AV. A Wi-Fi router is used to establish communication between components of our AV as well as reduce wire connection in our car. For example, the ROS client communicates with the ROS bridge server using a wireless connection via WebSocket protocol \cite{hussein2018ros}. We use a Raspberry Pi, a low-cost device, to read GPS location from Global Navigation Satellite System (GNSS). Subsequently, Raspberry Pi sends GPS data to a GPS node of ROS via wireless communication. Subsequently, we use a ROS bridge server to publish GPS data. This data is subscribed by a ROS bridge client running on different ROS computer.

\begin{figure}
\centerline{\includegraphics[width=0.45\textwidth]{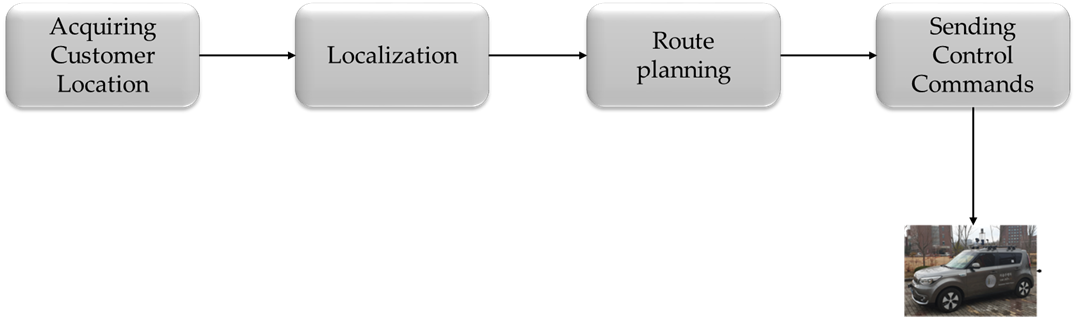}}
\caption{ROS environment and self-driving architecture of our autonomous car. We install Lidar, Odometry, inertial measurement unit (IMU) and GNSS in our KIA Soul EV car. ROS localization nodes are used to localize our car and search for customer location. Subsequently, ROS Route Planning node plans a travel route before sending control commands to our car using ROS Control Node.}
\label{fig13}
\end{figure}

ROS environment and self-driving architecture of our AV are shown in Fig. \ref{fig13}. Given the input location of an AV and a customer, we seek to search for an optimal route between the AV and the customer. We initiate the search from the AV location by a default radius (10 meters) and increase the radius if we do not find the target location of the customer. This localization task, research on Normal Distribution Transform (NDT) mapping \cite{mlvyeongmin}, is done through the Control and Planning loop.

\begin{figure*}
\centerline{\includegraphics[width=0.7\textwidth]{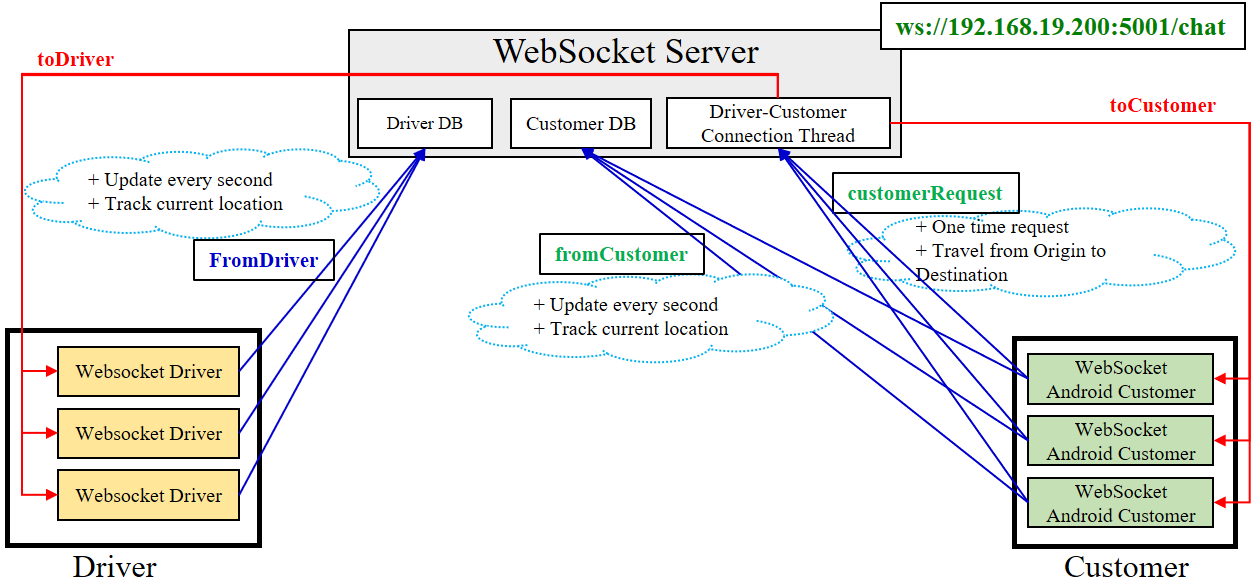}}
\caption{Booking system framework from network communication perspective. A pattern of our server WebSocket address is \textit{ws://IP\_address:port/chat}. The figure indicates that three have three drivers and three customers. The blue arrows are messages sent from customers and AVs in every second. It allows our server to track the current location of customers and AVs. The red arrows are messages dispatched from our server.}
\label{fig21}
\end{figure*}

\begin{figure*}
\centerline{\includegraphics[width=0.6\textwidth]{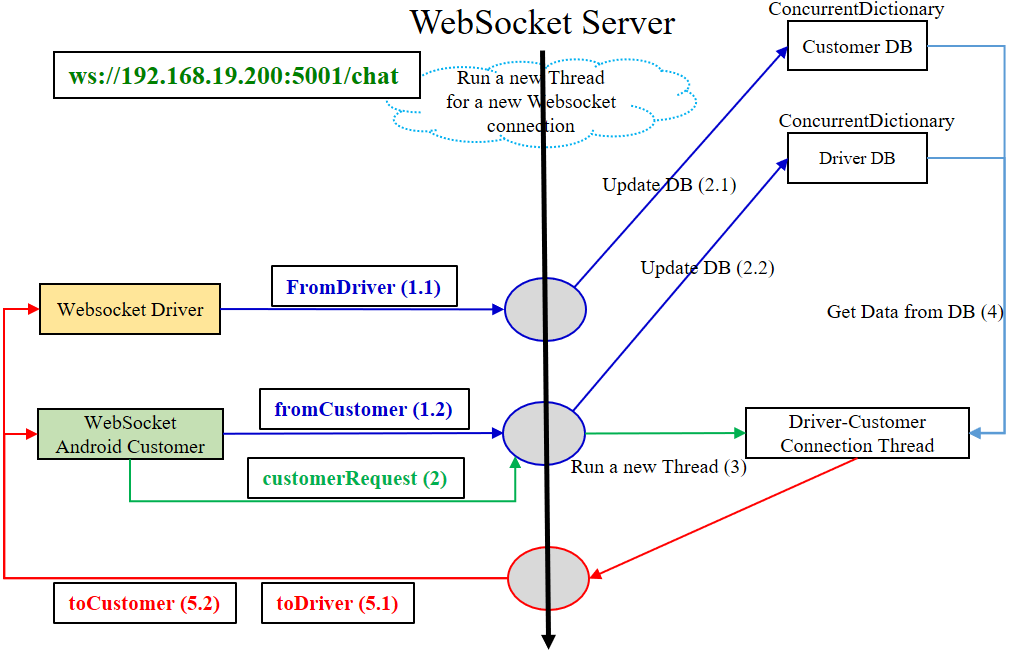}}
\caption{This figure decribes the workflow of our server application. Our application collects data from customers and AVs using JSON strings \{(1.1), (1.2)\}. It parses data from received strings and updates to concurrent dictionaries considered as tables on real-time database \{(2.1), (2.2)\}. If a customer sends a taxi request \{(2)\}, a new thread is created to get real-time data from database and sends back to our car \{(5.1)\} and the customer Android application \{(5.2)\}.}
\label{fig22}
\end{figure*}

The booking system framework from a network communication perspective is shown in Fig. \ref{fig21}. More detail, we open a WebSocket connection on a server using a public IP (Internet Protocol) address. This server listens to accept any WebSocket connection from customers and AVs. We create two separate tables on a real-time database (such as Firebase \cite{stonehem2016google}) to store the location of AVs and customers. Table \ref{tab1} describes the fields of the two tables.

The detail communication workflow is illustrated in Fig. \ref{fig22}. If an AV accepts a ride request from a customer, we create a scheduler to maintain a connection between an AV and a customer. That scheduler is called Driver-Customer connection thread. It handles communication between a pair of an AV and a customer.

\begin{table}
\caption{Real-time GPS location database}
\begin{center}
\begin{tabular}{|p{\mylength}|p{\mylength}|}
\hline
%\cline{2-4} 
\hspace{3.5em}\textbf{Customer Table} & \hspace{3.5em}\textbf{Driver Table}\\
\hline
(1) \textbf{ID}:  customer identification (unique)& (1) \textbf{ID}:  driver identification (unique)\\
(2) \textbf{Origin}: starting point of a ride (latitude, longitude)& (2) \textbf{CurrLocation}: current location of a driver (Latitude, Longitude)\\
(3) \textbf{Destination}: end point of a ride (latitude, longitude)& (3) \textbf{IsAvailable}: available status of a driver to serve customer\\
(4) \textbf{Stream}: an object represents a WebSocket connection between a customer and the server& (4) \textbf{Stream}: an object represents a WebSocket connection between a driver and  the server\\
\hline
\end{tabular}
\label{tab1}
\end{center}
\end{table}

In summary, our taxi booking system can be described as follows. Initially, AVs register all needed information shown in Driver table. Those data are stored in a server that supports communication between mobile applications and AVs. Firstly, a customer makes a taxi request containing its origin and destination. This request is sent directly to the server. Secondly, the server finds the best AV for a given customer request. The best AV might be the nearest or cheapest available AV. Thirdly, the server forwards the request message to the selected AV. If the AV accepts the request, we move to the fourth step. Fourthly, the server continuously sends the real-time location of the AV to the customer mobile apps. Concurrently, it also sends the real-time location of the customer application to the AV. Finally, the AV picks up the requested customer.

\section{Booking System Experiment}
\label{secIV}

\begin{figure*}
\centerline{\includegraphics[width=0.99\textwidth]{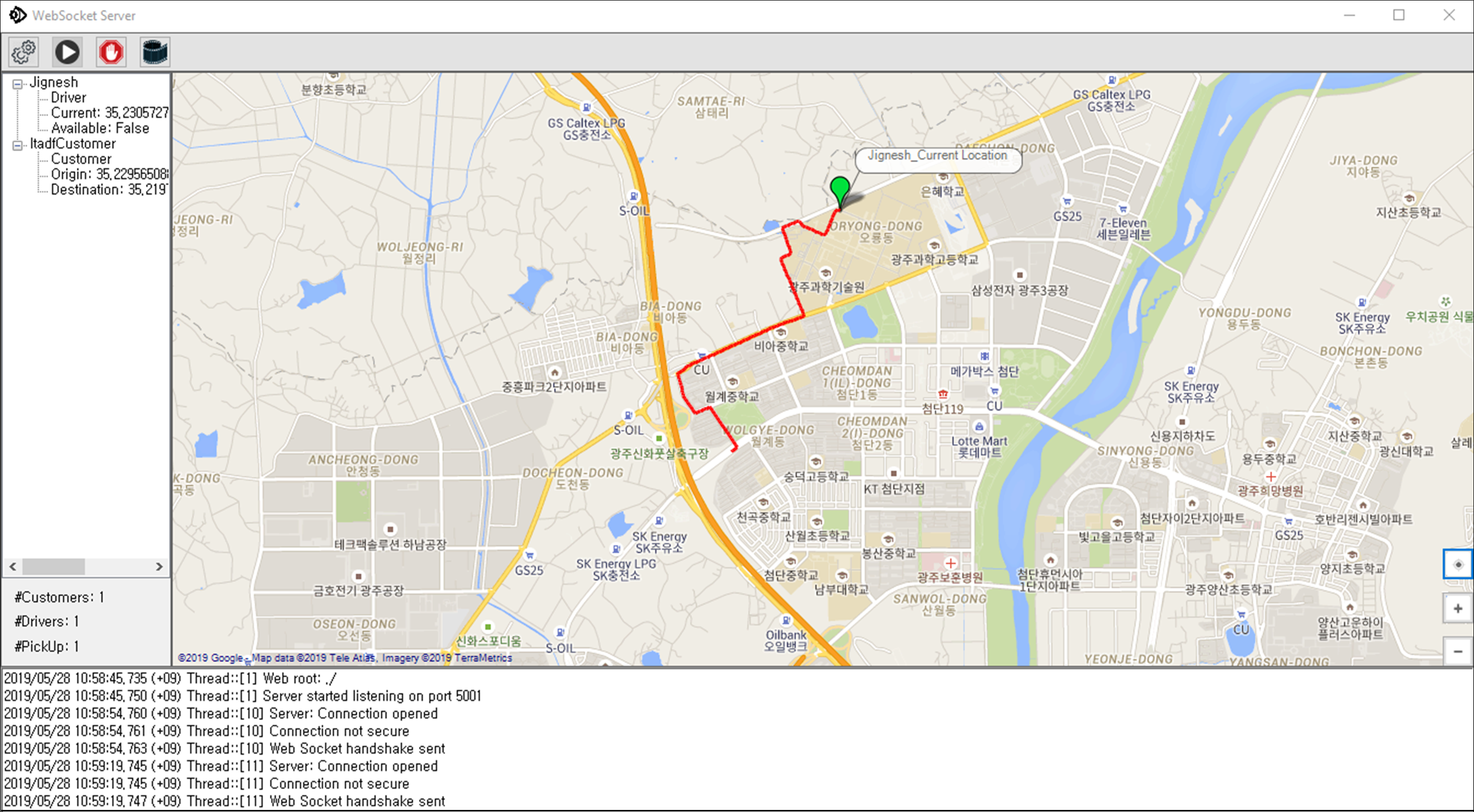}}
\caption{Our server application user interface runs on a Windows instance of AWS. It shows that three have one AV, one customer, and one pickup. AV is serving the customer with available status is \textit{false}. We can also see the current location of the AV as well as the origin and destination of the customer. The red line illustrates the result of tracking the AV (ID: \textbf{Jignesh}).}
\label{fig31}
\end{figure*}

\begin{figure*}
\centerline{\includegraphics[width=0.96\textwidth]{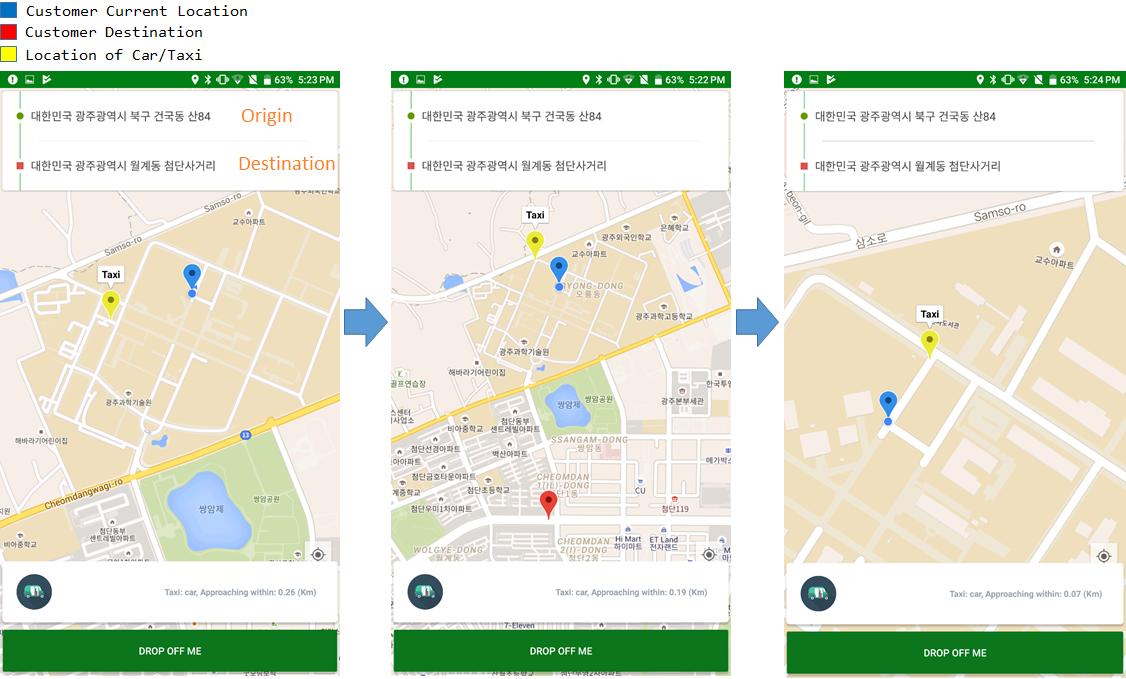}}
\caption{our car (yellow marker) moves on Google map of our Android application. The red marker indicates the destination chosen by the customer. A layout above "\textit{Drop Off Me}" button displays the distance between the customer and an our AV. This distance is updated if the location of our car or customer changes. The origin and destination are written in Korean.}
\label{fig32}
\end{figure*}

We use KIA Soul EV and install additional equipment, such as Lidar, Odometry, IMU, GNSS, and cameras to make our own self-driving car.  The detail of this work is thoroughly presented in our previous works\cite{munir2018autonomous,mlvazam,mlvyeongmin,mlvaiom,azam2018object}. We also build our own mobile application and the server application which is installed on a Windows instance of AWS. We subscribe to KT (Korea Telecom), which is a telephone company in South Korea, to access the Internet from our car using a mobile KT egg Wi-Fi device.

Fig. \ref{fig31} shows our server user interface running on Windows. It can display driver and customer information read from the real-time database. For example, it shows a number of customers, drivers, and pickup done by all drivers. We can also know the status (available or not) as well as the location of drivers and customers. Furthermore, we can track the movement of a user (driver or customer) on Google map.

The application allows customers to choose the origin and destination of a ride. This information is updated in Customer table. For example, a customer uses our android application to send a booking car request to our server. Afterwards, the application updates its current location to the server whenever GPS location of the customer is changed by the following JSON (JavaScript Object Notation) format string: \textit{\{'ID':'Itadf', Type':'Customer', Origin':'35.228683, 126.844866'\}}. The string specifies that there has a customer with an ID (\textbf{Itadf}) from latitude \textit{35.228683} and longitude \textit{126.844866}. If a customer requests a ride, it sends the following JSON string: \textit{\{'ID':'Itadf', 'Type':'Request', 'Destination':'35.227139, 126.838194'\}}. If an AV accepts the request, the Android application will receive a JSON message from the server every second. 

Our autonomous car periodically updates its current location to the server and waits for a booking request from a customer. The car is responsible for publishing his current location by the following JSON string: \textit{\{'ID':'Jignesh', Type':'Driver', Location':'35.228683, 126.844866'\}}. This information is updated in Driver table.

Once a customer requests a ride, it will receive a JSON string as follows \textit{\{'Customer':'Itadf', 'Origin':'35.228683, 126.844866', 'Destination':'35.227139, 126.838194'\}}. An AV can update its available status as follows \textit{\{'ID':'Jignesh', 'Type':'Driver', 'Location':'35.228683, 126.844866', 'Available':‘True'\}}. This JSON is only sent once if the driver is available. A sequence of images from left to right of Fig. \ref{fig32} illustrates our car (yellow marker) is approaching customer location (blue marker).

\begin{figure}
\centerline{\includegraphics[width=0.48\textwidth]{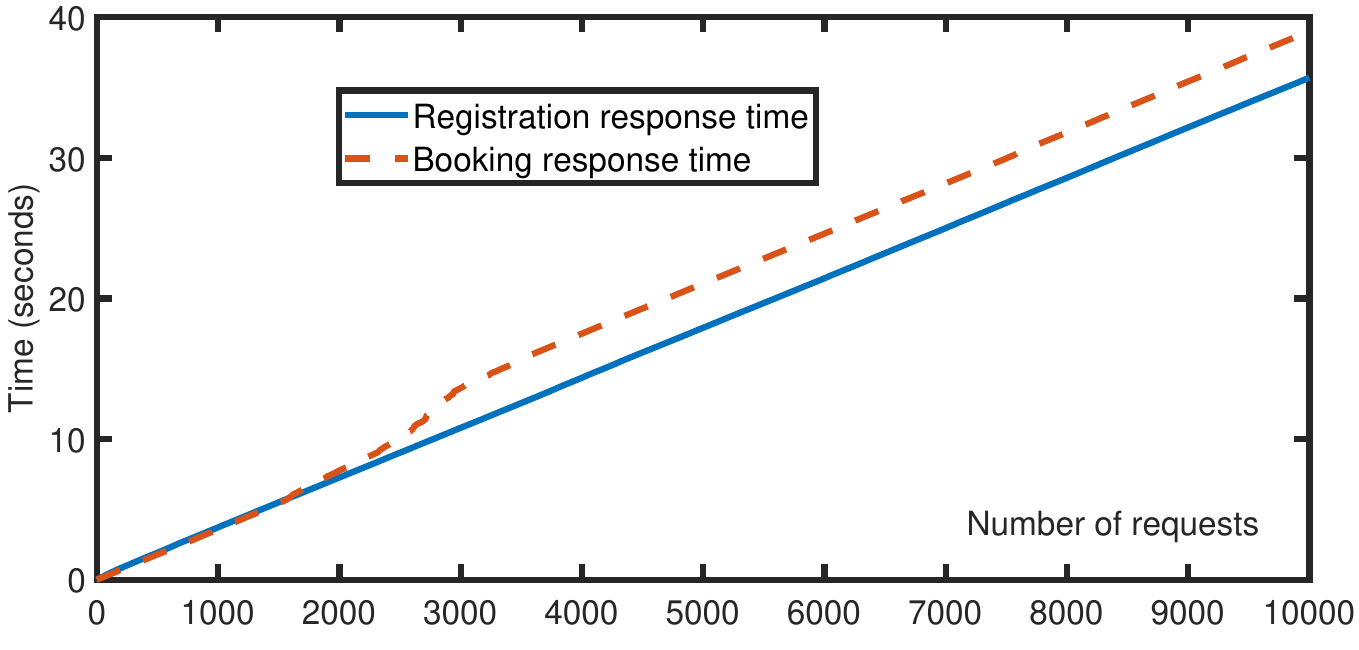}}
\caption{Our booking system performance in term of response time.}
\label{figtime}
\end{figure}

We deploy our booking system on a compute Core(TM) i5\-8500 CPU@3.00GHz. We set up virtual clients (drivers and customers) that send simultaneously a number of registration or booking requests to our system. Fig. \ref{figtime} describes the registration and booking response time of our system. Registration response time increases linearly as a number of requests increases. Booking response time is slightly higher than registration because booking requests require more operations on finding the best-matched taxi.
%(response time equals a multiplication of 0.04 times a number of requests)

\section{Conclusion}
\label{secV}
We presented a research on taxi booking system for autonomous vehicles. The system can operate automatically without any support from human, and customers can feel more present in a comfortable ride. We expect that this technology can be deployed in the near future. Machine learning and distributed system techniques can be the future research topics for automating taxi booking operation. Given a customer request (either booking via a phone call or mobile application), machine learning algorithms can optimally search for the best AV taxi based on learning data collected from a distributed booking system. As addressed by the authors in \cite{freedman2018autonomous,dandl2018comparing}, an AV taxi service will potentially safer and be less expensive than a human-driven taxi system. However, this automatic system poses many security problems which are an open research area. For example, a hacker can control the system which could lead to fatal injury accidents.

\section*{Acknowledgment}

This work was supported by  GIST Research Institute(GRI) grant funded by the GIST in 2019, and by Institute of Information \& Communications Technology Planning \& Evaluation (IITP) grant funded by the Korea government (MSIT)(No.2014-3-00077, AI National Strategy Project).

\bibliography{ref}
\bibliographystyle{IEEEtran}

\end{document}